\documentclass[pdftex,twocolumn,epjc3]{svjour3}
\usepackage{amsmath}
\usepackage{amssymb}

\RequirePackage[T1]{fontenc}
\smartqed
\RequirePackage{graphicx}
\RequirePackage{mathptmx}
\RequirePackage{flushend}
\RequirePackage[numbers,sort&compress]{natbib}
\RequirePackage[colorlinks,citecolor=blue,urlcolor=blue,linkcolor=blue]{hyperref}

\journalname{Eur. Phys. J. C}

\begin{document}

\title{On Aharonov-Casher bound states}

\author{E. O. Silva\thanksref{e1,addr1}
        \and
        F. M. Andrade\thanksref{e2,addr2}
        \and
        C. Filgueiras\thanksref{e3,addr3}
        \and
        H. Belich\thanksref{e4,addr4}
}

\thankstext{e1}{e-mail: edilbertoo@gmail.com}
\thankstext{e2}{e-mail: fmandrade@uepg.br}
\thankstext{e3}{e-mail: cleversonfilgueiras@yahoo.com.br}
\thankstext{e4}{e-mail: belichjr@gmail.com}

\institute{Departamento de F\'{i}sica,
           Universidade Federal do Maranh\~{a}o,
           Campus Universit\'{a}rio do Bacanga,
           65085-580 S\~{a}o Lu\'{i}s-MA, Brazil\label{addr1}
           \and
           Departamento de Matem\'{a}tica e Estat\'{i}stica,
           Universidade Estadual de Ponta Grossa,
           84030-900 Ponta Grossa-PR, Brazi\label{addr2}
           \and
           Departamento de F\'{\i}sica,
           Universidade Federal de Campina Grande,
           Caixa Postal 10071,
           58109-970 Campina Grande-PB, Brazil\label{addr3}
           \and
           Departamento de F\'{i}sica e Qu\'{i}mica,
           Universidade Federal do Esp\'{i}rito Santo,
           29060-900, Vit\'{o}ria-ES, Brazil\label{addr4}
         }

\date{Received: date / Accepted: date}

\maketitle

\begin{abstract}
In this work bound states for the Aharonov-Casher problem are
considered.
According to Hagen's work on the exact equivalence between
spin-1/2 Aharonov-Bohm and Aharonov-Casher effects, is known
that the $\boldsymbol{\nabla}\cdot\mathbf{E}$ term cannot be
neglected in the Hamiltonian  if the spin of particle is
considered.
This term leads to the existence of a singular potential at the
origin.
By modeling the problem by boundary conditions at the origin
which arises by the self-adjoint extension of the Hamiltonian,
we derive for the first time an expression for the bound state
energy of the Aharonov-Casher problem.
As an application, we consider the Aharonov-Casher plus a
two-dimensional harmonic oscillator.
We derive the expression for the harmonic oscillator energies
and compare it with the expression obtained in the case without
singularity.
At the end, an approach for determination of the self-adjoint
extension parameter is given. In our approach, the parameter is
obtained essentially in terms of physics of the problem.
\end{abstract}

\section{Introduction}
\label{sec:introduction}

Physical processes which exhibit a cyclic evolution play an
important role in the description of quantum systems in a
periodically changing environment.
The system can present a classical description, such as a
magnetic dipole precession around an external magnetic field, or
quantum behavior such as the electron in a condensate, produced
by collective motion of atoms.
In this context, the role played by the electromagnetic vector
potential is remarkable.
The presence of a vector potential in a region where it does not
produces an electric or a magnetic field in the configuration
space of free electrons, can influence the interference pattern.
Considering a charged particle which propagates in
a region with no external magnetic field (force-free region), it
is verified that the corresponding wave function may develop a
quantum phase:
$\langle b|a\rangle_{inA}=\langle b|a\rangle_{A=0}
\left\{\exp(iq\int_{a}^{b}\mathbf{A}\cdot d\mathbf{l})\right\}$,
which describes the real behavior of the electrons propagation.

Topological effects in quantum mechanics are  phenomena that
present no classical counterparts, being associated with
physical systems defined on a multiply connected space-time
\cite{PRL.1995.74.2847}.
This issue has received considerable attention since the
pioneering work by Aharonov and Bohm \cite{PR.1959.115.485},
where they demonstrated that the vector potential may induce
measurable physical quantum phases even in a force-free
region, which constitutes the essence of a topological
effect.
The induced phase does not depend on the specific path
described by the particle nor on its velocity
(non-dispersiveness).
Instead, it is intrinsically related to the non-simply
connected nature of the space-time and to the associated
winding number.
This is a first example of generation of a topological phase,
the so called Aharonov-Bohm (AB) effect.
Many years later, Aharonov and Casher \cite{PRL.1984.53.319}
argued that a quantum phase also appears in the wave
function of a spin-1/2 neutral particle with anomalous
magnetic moment, $\mathbf{\mu}$, subject to an electric
field arising from a charged wire.
This is the well-known Aharonov-Casher (AC) effect, which is
related to the AB effect by a duality operation
\cite{PRL.1990.64.2347}.
This phase is well established if the wire penetrates
perpendicularly to the plane of the neutral particle
motion.
The  AC phase and analogous effects have been studied in
several branches of physics in recent years (see for example
the Refs.
\cite{PRD.2011.83.125025,PRA.2007.76.012113,PRD.2009.80.024033,
IJMPD.2010.19.85,AdP.2011.523.910,JMP.2011.52.063505,
JHEP.2004.2004.16,PRL.1989.62.1071,JMP.2012.53.023514,
AdP.2010.522.447,PRD.2010.82.084025,PLA.2010.374.3143,
PRA.2009.80.032106,PRD.2009.80.024033}).

In order to study the quantum dynamics of these systems we must consider
the full Hamiltonian (including the electromagnetic field sources).
For example, when we are studying the nonrelativistic dynamics of a
spin-1/2 charged particle in an electromagnetic field
\footnote{In particular electric (magnetic) field generated by an
  infinitely long, infinitesimally thin line of charge (or solenoids)
  which we are interested here.},
it is common to neglect terms that explicitly depend
on the spin, namely, $\boldsymbol{\nabla}\times\mathbf{A}$ and
$\boldsymbol{\nabla}\cdot \mathbf{E}$ terms.
However, these sources lead to singular solutions at the origin
\cite{AoP.2010.325.2529,AoP.2008.323.3150,AoP.1996.251.45,
PRD.1994.50.7715,JSP.2008.133.1175,JPA.2010.43.354011}.
Hagen \cite{PRL.1990.64.2347} showed that there is an exact
equivalence between the AB and AC effects when comparing the
$\boldsymbol{\nabla}\cdot\mathbf{E}$ term in the AC Hamiltonian
with the $\boldsymbol{\nabla}\times\mathbf{A}$ term in the AB
Hamiltonian.
He concluded that the  $\boldsymbol{\nabla}\cdot\mathbf{E}$ term
cannot be neglected when we consider the spin of the particle
\cite{PRL.1990.64.2347}.
Recently, the $\boldsymbol{\nabla}\cdot \mathbf{E}$ term was
considered by Shikakhwa \textit{et. al}
\cite{JPA.2010.43.354008} in the scattering scenario.
However, an analysis of the energy spectrum considering this
term is not found in the literature.
This is the main subject of present work, which aims to study the AC
bound states.

In AB-like systems, an interesting question which emerges nowadays is the
study of the phase generation in real situations, such as in condensed
matter, putting, for example, a wire in some material
\cite{Book.2010.Ihn}. 
To reproduce this scenario it is advisable to couple
the charged wire to a harmonic oscillator (HO), and then analyze this
influence on the  dynamics of the neutral particle. The vibration of a
crystalline lattice, in which the wire is embedded, is simulated by
HO behavior \cite{PRB.1994.50.8460,PLA.1992.166.377}.

In this paper, we analyze the AC problem taking into account the
$\nabla \cdot E=\lambda \delta (r)/2\pi r$ term in the nonrelativistic
Hamiltonian. Using the self-adjoint extension method
\cite{Book.1975.Reed.II}, we model the problem by boundary conditions
\cite{CMP.1991.139.103} and determine the expression for the energy
spectrum of the particle and compare with the case where one neglects
the such term.
We  also address the AC problem interacting with a two-dimensional
harmonic oscillator located at the origin of a polar coordinate system
and compare our results with those known in the literature.
In Sec. \ref{sec:eqmotion} the equation of motion for the AC problem is
derived in the nonrelativistic limit.
In Sec. \ref{sec:bstates_AC} the bound states for AC problem is
examined.
Expressions for the wave functions and bound state energies are obtained
fixing the physical problem in the $r=0$ region, without any arbitrary
parameter.
In Sec. \ref{sec:acho} the AC problem interacting with a two-dimensional
harmonic oscillator located at the origin of a polar coordinate system
is considered.
Again, modeling the problem by boundary conditions, we found the
expression for the energy spectrum of the oscillator in terms of the
physics of the problem without any arbitrary parameter.
In Sec. \ref{sec:dsaep} we present the procedure to determine the
self-adjoint extension parameter. In Sec. \ref{sec:conclusions} a brief
conclusion is given.

\section{The equation of motion for the AC problem}
\label{sec:eqmotion}

In order to study the dynamics of a neutral particle with a magnetic
moment $\mu$ in a flat space-time we start with the Dirac equation
(with $\hbar=c=1$)
\begin{equation}
  \left[
    i\gamma^{\mu}\partial_{\mu}-\frac{\mu}{2}
    \sigma^{\mu \nu}F_{\mu \nu}-M
  \right]\psi=0, \qquad \mu,\nu=0,1,2,
\label{eq:edirac}
\end{equation}
where $M$ is the mass of the particle and $\psi$ is a four-component
spinorial wave function and the $\gamma$-matrices obey the commutator
relation
\begin{equation}
  \sigma^{\mu \nu}=\frac{i}{2}
  \left[
    \gamma^{\mu},\gamma^{\nu}
  \right] .
\end{equation}
By introducing a spin projection parameter $s$ in 2+1 dimensions
Eq. \eqref{eq:edirac} reduces to a set of two two-component wave
functions.
In this case, the Dirac matrices are conveniently defined in the Pauli
representation \cite{PRL.1990.64.2347}
\begin{equation}
  \beta =\gamma^{0}= \sigma_{3},\qquad
  \beta  \gamma^{1}= \sigma_{1},\qquad
  \beta  \gamma^{2}=s\sigma_{2}.
\end{equation}
In this representation, the Dirac equation is found to be
\cite{PRL.1990.64.2347}
\begin{equation}
\left[
  M\beta +\beta \boldsymbol{\gamma}\cdot
  \left( \frac{1}{i}\mathbf{{\nabla}}-
    \mu s\mathbf{E}'
  \right)
\right]
\psi =\bar{\mathcal{E}}\psi,\qquad(i=1,2)
 \label{eq:deq}
\end{equation}
where $E'_{i}\equiv \epsilon _{ij}E_{j}$ denotes the
dual field and $\epsilon_{ij}=-\epsilon_{ji}$, $\epsilon_{12}=+1$.
By applying the matrix operator
\begin{equation}
  \left[
    M+\beta \bar{\mathcal{E}}-\beta\boldsymbol{\gamma}\cdot
    \left(
      \frac{1}{i}\boldsymbol{\nabla}-
      \mu s\mathbf{E}'
    \right)
  \right] \beta
\end{equation}
in Eq. \eqref{eq:deq} we obtain
\begin{align}
  \left(
    \bar{\mathcal{E}}^{2}-M^{2}\right) \psi
  ={} & -
  \left(\boldsymbol{\gamma}\cdot\boldsymbol{\pi}\right)
  \left(\boldsymbol{\gamma}\cdot\boldsymbol{\pi}\right)
  \psi  \nonumber \\
  ={} &
  \left[
    \boldsymbol{\pi }^{2}+
    \mu \sigma _{3}\left(\boldsymbol{\nabla}\cdot
      \mathbf{E}'\right)
  \right] \psi,
  \label{eq:hrel}
\end{align}
where
$\boldsymbol{\pi}=\frac{1}{i}\boldsymbol{\nabla}-
\mu s\mathbf{E}'$.

In the usual AC effect the field configuration (in cylindrical
coordinates) is given by
\begin{equation}
  \mathbf{E}=\frac{\lambda}{2\pi\varepsilon_{0}}
  \frac{\mathbf{\hat{r}}}{r},\qquad
  \boldsymbol{\nabla}\cdot\mathbf{E}=
  \frac{\lambda}{2\pi\varepsilon_{0}}\frac{\delta(r)}{r},
  \label{eq:acconf}
\end{equation}
where $\mathbf{E}$, is the electric field generated by an infinite
charge filament and $\lambda $ is the charge density along the $z$-axis.
To analyze the nonrelativistic limit we assume
\begin{align}
  \label{eq:glim}
  \mathcal{\bar{E}}  = {} &M+\mathcal{E},  \nonumber \\
  \mathcal{E}        \ll {}&M,
\end{align}
and Eq. \eqref{eq:hrel}, using Eq. \eqref{eq:acconf}, assumes the form
\begin{equation}
  \hat{H}_{NR}\psi =\mathcal{E}\psi ,
  \label{eq:hnr}
\end{equation}
where
\begin{equation}
  \hat{H}_{NR}=\frac{1}{2M}
  \left[
    \frac{1}{i}\boldsymbol{\nabla}+
    s\eta\frac{\boldsymbol{\hat{\varphi}}}{r}
  \right]^{2}+
  \frac{\eta \sigma_{3}}{2M}\frac{\delta(r)}{r},
  \label{eq:hamil2d}
\end{equation}
where
\begin{equation}
\eta=\frac{\mu\lambda}{2\pi\varepsilon_{0}}.
\end{equation}
The nonrelativistic Hamiltonian above describes the planar
dynamics of a spin-1/2 neutral particle with a magnetic moment
$\mu$ in an electric field.

Before we go on to a calculation of the bound states some remarks
on Hamiltonian in \eqref{eq:hamil2d} are in order.
If we do not take into account the $\boldsymbol{\nabla}\cdot\mathbf{E}$
term, the resulting Hamiltonian, in this case, is essentially
self-adjoint and positive definite \cite{PRA.2002.66.032118}.
Therefore, its spectrum is $\mathbb{R}^{+}$, it is translationally
invariant and there are no bound states
The introduction of $\boldsymbol{\nabla}\cdot\mathbf{E}$ changes
the situation completely.
The singularity at origin due the
$\boldsymbol{\nabla}\cdot\mathbf{E}$, is physically equivalent
to extracting this single point from the plane $\mathbb{R}^2$ and
in this case the translational invariance is lost together with
the self-adjointness.
This fact has impressive consequences in the spectrum of the
system  \cite{Book.2008.Oliveira}.
Since we are effectively excluding a portion of space accessible
to the particle we must guarantee that the Hamiltonian is
self-adjoint in the region of the motion, as is necessary for
the generator of time evolution of the wave function.
The most adequate approach for studying this scenario is the
theory of self-adjoint extension of symmetrical operators of von
Neumann-Krein
\cite{Book.2004.Albeverio,Book.1975.Reed.II,Book.1993.Akhiezer}.
The existence of a negative eigenvalue in the spectrum can be
considered rather unexpected, since the actions of it suggest it is
positive definite operator.
However, the positivity of such an operator just not only depends on
its action, but also depends on its domain.
Moreover, the $\boldsymbol{\nabla}\cdot\mathbf{E}$ gives rising
to a two-dimensional $\delta$ function potential at origin, and it is a
well known fact that an attractive $\delta$ function allows at least
one bound state \cite{Book.2005.Griffiths,Book.2004.Albeverio}.

In particular, we analyze the changes in the energy levels and wave
functions of the particle in the region near the charge filament.
For this system, the commutator
\begin{equation}
\left[ \hat{H}_{NR},\hat{J}_{z}\right] =0,
\end{equation}
where $\hat{J}_{z}=-i\partial_{\varphi}+\sigma_{z}/2$ is the
total angular momentum operator in the $z$ direction.
So, the solution for the Schr\"{o}dinger equation
\eqref{eq:hnr} can be written in the form
\begin{equation}
  \Phi(r,\varphi)=
  \left[
    \begin{array}{c}
      f_{\mathcal{E}}(r)e^{i(m_{j}-1/2)\varphi} \\
      g_{\mathcal{E}}(r)e^{i(m_{j}+1/2)\varphi}
    \end{array}
  \right],
  \label{eq:eqref}
\end{equation}
with $m_{j}=m+1/2=\pm 1/2,\pm 3/2,\ldots$, $m\in\mathbb{Z}$.
By substitution Eq. \eqref{eq:eqref}
into Eq. \eqref{eq:hnr}, the radial equation for $f_{\mathcal{E}}(r)$
becomes
\begin{equation}
Hf_{\mathcal{E}}(r)=\mathcal{E}f_{\mathcal{E}}(r),  \label{eq:eigen}
\end{equation}
where
\begin{equation}
H=
H_{0}+
U_{\rm short},
\label{eq:hfull}
\end{equation}
\begin{equation}
  H_{0}=-\frac{1}{2M}
  \left[
    \frac{d^{2}}{dr^{2}}+\frac{1}{r}\frac{d}{dr}-\frac{\xi^{2}}{r^{2}},
  \right],
  \label{eq:hzero}
\end{equation}
\begin{equation}
U_{\rm short}=\frac{\eta}{2M}\frac{\delta(r)}{r},
\label{eq:ushort}
\end{equation}
with
\begin{equation}
  \xi =m+s\eta.
\label{eq:leff}
\end{equation}
In what follows we assume that $\eta<0$ to ensure we have an attractive
$\delta$ function with at least one bound state.

Since we are considering
an infinite hollow cylinder of radius $r_{0}$  with charge density per
unit length $\lambda$, is suitable to rewrite the short-range potential
as (see \cite{PRD.2012.85.041701} and references therein)
\begin{equation}
\label{eq:ushort_r0}
\overline{U}_{\rm short}(r)=\frac{\eta}{2M}
\frac{\delta(r-r_{0})}{r_{0}}.
\end{equation}
Although the functional structures of  $U_{\rm short}$ and
$\overline{U}_{\rm short}$ are quite different, as
discussed in \cite{PRL.1990.64.503}, we are free to use any form of
potential provided that only the contribution of the form
\eqref{eq:ushort}  is excluded.

\section{Bound states for AC problem}
\label{sec:bstates_AC}

Now, the goal is to find the bound states for the Hamiltonian
\eqref{eq:hfull}.
This Hamiltonian includes a short-range interaction
$\overline{U}_{\rm short}$ modeled by a $\delta$ function.
Such kind of point interaction also appears in several
AB-like problems, e.g., AB problem of spin-1/2 particles
\cite{Book.1988.Demkov,PRD.1994.50.7715,PRD.2012.85.041701,PLB.2013.719.467},
AC problem in a CPT-odd Lorentz-violating background
\cite{EPL.2013.101.51005,arXiv:1303.1660}, in
the coupling between wave functions and conical defects/cosmic strings
\cite{JMP.2012.53.122106,PRD.1996.53.6829}, and coupling between wave
functions and torsion \cite{PRD.2009.80.024033}.

To deal with the singularity point at $r=0$, we follow the approach in
\cite{CMP.1991.139.103}. Then we temporarily forget the
$\delta$ function potential and find which boundary conditions are
allowed for $H_{0}$. This is the scope of the self-adjoint
extension, which consists in determining the complete domain of an
operator, i.e., its complete set of wave functions. But the self-adjoint
extension provides us with an infinity of possible boundary conditions
and therefore it cannot give us the true physics of the
problem. 
Nevertheless, once having fixed the physics at $r=0$
\cite{PRD.1989.40.1346,NPB.1989.328.140}, we are able to fit any
arbitrary parameter coming from the self-adjoint extension and then we
have a complete description of the problem.

Since we have a singular point, even if
$H_{0}^{\dagger}=H_{0}$, we must guarantee that the
Hamiltonian is self-adjoint in the region of motion for their domains
might be different. The von Neumann-Krein method
\cite{Book.1975.Reed.II} is used to find the self-adjoint extensions. An
operator $H_{0}$ with domain $\mathcal{D}(H_{0})$ is
self-adjoint if
$\mathcal{D}(H_{0}^{\dagger})=\mathcal{D}(H_{0})$
and $H_{0}^{\dagger}=H_{0}$. In order to proceed
with the self-adjoint extension, we must find the deficiency
subspaces $\mathcal{N}_{\pm}$, with dimensions $n_{+}$ and $ n_{-}$,
which are called deficiency indices of $H_{0}$. A necessary
and sufficient condition for $H_{0}$ to be self-adjoint is
that $n_{+}=n_{-}=0$. On the other hand, if $n_{+}=n_{-}\geq 1$ then
$H_{0}$ has an infinite number of self-adjoint extensions
parametrized by a unitary $n\times n$ matrix, where $n=n_{+}=n_{-}$.

The potential in this case is purely radial and we decompose the Hilbert
space $\mathfrak{H}=L^{2}(\mathbb{R}^{2})$ with respect to the angular
momentum $\mathfrak{H}=\mathfrak{H}_{r}\otimes\mathfrak{H}_{\varphi}$, where
$\mathfrak{H}_{r}=L^{2}(\mathbb{R}^{+},rdr)$ and
$\mathfrak{H}_{\varphi}=L^{2}({S}^{1},d\varphi)$, with
${S}^{1}$ denoting the unit sphere in $\mathbb{R}^{2}$.
The operator $-{\partial^{2}}/{\partial\varphi^{2}}$ is essentially
self-adjoint in $\mathfrak{H}_{\varphi}$
\cite{Book.1975.Reed.II} and we obtain the operator $H_{0}$ in
each angular momentum sector. We have to pay special attention for the
radial eigenfunctions, due to the singularity at $r=0$.

Next, we substitute the problem in Eq. \eqref{eq:eigen} by
\begin{equation}
H_{0}f_{\varrho,\mathcal{E}}=\mathcal{E}f_{\varrho,\mathcal{E}},
\label{eq:hzeroeigen}
\end{equation}
with $f_{\varrho,\mathcal{E}}$ labeled by a parameter $\varrho$
which is related to the behavior of the wave function in the limit
$r\rightarrow r_{0}$. But we cannot impose any boundary condition
(e.g. $f=0$ at $r=0$) without discovering which boundary conditions
are allowed to $H_{0}$.

In order to find the full domain of $H_{0}$ in
$L^{2}(\mathbb{R}^{+},rdr)$, we have to find its deficiency subspaces.
To do this, we solve the eigenvalue equation
\begin{equation}
H_{0}^{\dagger}f_{\pm}=\pm i k_{0} f_{\pm},
\label{eq:eigendefs}
\end{equation}
where $H_{0}^{\dagger}$ is given by Eq. \eqref{eq:hzero} and
$k_{0}\in\mathbb{R}$ is introduced for dimensional reasons.
The only square-integrable functions which are solutions of
Eq. \eqref{eq:eigendefs} are the modified Bessel functions
\begin{equation}
  f_{\pm}(r)={\rm const.}\; K_{\xi}(r\sqrt{\mp \varepsilon}),
\end{equation}
with $\varepsilon=2 i M k_{0}$. These functions are square-integrable
only in the range $\xi\in(-1,1)$, for which $H_{0}$ is not
self-adjoint, and the dimension of such deficiency subspaces is
$(n_{+},n_{-})=(1,1)$.
So we have two situations for $\xi$, i.e.,
\begin{eqnarray}
\label{eq:xirange}
-1<\xi<0,\\ \nonumber
0<\xi<1.
\end{eqnarray}
To treat both cases of Eq. \eqref{eq:xirange} simultaneously,
it is  more convenient to use
\begin{equation}
  f_{\pm}(r)={\rm const.}\; K_{|\xi|}(r\sqrt{\mp \varepsilon}).
\end{equation}
Thus, the domain
$\mathcal{D}(H_{0,\varrho})$ in $L^{2}(\mathbb{R}^{+},rdr)$ is
given by the set of functions
\cite{Book.1975.Reed.II}
\begin{equation}
 \label{eq:domain}
  f_{\varrho,\mathcal{E}}(r)=f_{|\xi|}(r)+
  C\left[
    K_{|\xi|}(r\sqrt{-\varepsilon})+
    e^{i\varrho}K_{|\xi|}(r\sqrt{\varepsilon})
\right] ,
\end{equation}
where $f_{|\xi|}(r)$, with
$f_{|\xi|}(r_{0})=\dot{f}_{|\xi|}(r_{0})=0$ ($\dot{f}\equiv df/dr$),
is the regular wave function when we do not have
$\overline{U}_{\rm short}(r)$. The last term in
Eq. \eqref{eq:domain} gives the correct behavior for the wave function
when $r=r_{0}$. The parameter $\varrho \in [0,2\pi)$ represents a
choice for the boundary condition in the region around $r=r_{0}$ and
describes the coupling between
$\overline{U}_{\rm short}(r)$ and the wave function. As we
shall see below, the physics of the problem determines such parameter
without ambiguity.

To find a fitting for $\varrho$ compatible with
$\overline{U}_{\rm short}(r)$, we write
Eqs. \eqref{eq:eigen} and \eqref{eq:hzeroeigen} for $\mathcal{E}=0$
\cite{CMP.1991.139.103},
\begin{equation}
  \left[
    \frac{d^{2}}{dr^{2}}+\frac{1}{r}\frac{d}{dr}-\frac{\xi^{2}}{r^{2}}+
    \overline{U}_{\rm short}
  \right] f_{0}=0,
\label{eq:truestatic}
\end{equation}
\begin{equation}
  \left[
    \frac{d^{2}}{dr^{2}}+\frac{1}{r}\frac{d}{dr}-\frac{\xi^{2}}{r^{2}}
  \right] f_{\varrho,0}=0,
\label{eq:rhostatic}
\end{equation}
implying the zero-energy solutions $f_{0}$ and $f_{\varrho,0}$,
and  we require the continuity for the logarithmic
derivative \cite{CMP.1991.139.103}
\begin{equation}
r_{0}\frac{\dot{f}_{0}(r_{0})}{f_{0}(r_{0})}=
r_{0}\frac{\dot{f}_{\varrho,0}(r_{0})}{f_{\varrho,0}(r_{0})}.
\label{eq:ftrue}
\end{equation}
The left-hand side of Eq. \eqref{eq:ftrue} is achieved integrating
Eq. \eqref{eq:truestatic} from $0$ to $r_{0}$,
\begin{align}
  \label{eq:inth0}
  \int_{0}^{r_{0}}\frac{1}{r}\frac{d}{dr}
  (r\frac{df_{0}(r)}{dr}) rdr
  = {} & \eta \int_{0}^{r_{0}}f_{0}(r)\frac{\delta(r-r_{0})}{r}
  rdr \nonumber \\
  {} & +\xi^{2}\int_{0}^{r_{0}}\frac{f_{0}(r)}{r^{2}} rdr.
\end{align}
From \eqref{eq:truestatic}, the behavior of $f_{0}$ as $r\to 0$
is $f_{0}\sim r^{|\xi|}$, so we find
\begin{equation}
\int_{0}^{r_{0}}\frac{f_{0}(r)}{r^{2}}rdr\approx
\int_{0}^{r_{0}}r^{|\xi|-1}dr \to 0.
\end{equation}
So, we arrive at
\begin{equation}
  r_{0}\frac{\dot{f}_{0}(r_{0})}{f_{0}(r_{0})}\approx\eta.
  \label{eq:ftrueleft}
\end{equation}
The right-hand side of Eq. \eqref{eq:ftrue} is calculated using
the asymptotic representation for the modified Bessel functions
in the limit $z\rightarrow 0$,
\begin{equation}
  \label{eq:besselasympt}
  K_{\nu}(z) \sim
  \frac{\pi}{2\sin(\pi \nu )}
  \left[
    \frac{z^{-\nu}}{2^{-\nu}\Gamma(1-\nu)}-
    \frac{z^{ \nu}}{2^{ \nu}\Gamma(1+\nu)}
  \right],
\end{equation}
in Eq. \eqref{eq:domain} and it takes the form
\begin{equation}
  r_{0}\frac{\dot{f}_{\varrho,0}(r_{0})}{f_{\varrho,0}(r_{0})}=
  \frac{\dot{\Omega}_{\varrho}(r_{0})}{\Omega_{\varrho}(r_{0})},
  \label{eq:ftrueright}
\end{equation}
where
\begin{align}
  \label{eq:omegas}
  \Omega_{\varrho}(r)=& {}
  \left[
    \frac{(r \sqrt{-\varepsilon})^{-|\xi|}}
    {2^{-|\xi|}\Gamma(1-|\xi|)} -
    \frac{(r \sqrt{-\varepsilon})^{|\xi|}}
    {2^{|\xi|}\Gamma(1+|\xi|)}
  \right]
  \nonumber \\
  & {}
  + e^{i\varrho}
  \left[
    \frac{(r \sqrt{\varepsilon})^{-|\xi|}}
    {2^{-|\xi|}\Gamma(1-|\xi|)} -
    \frac{(r \sqrt{\varepsilon})^{|\xi|}}
    {2^{|\xi|}\Gamma(1+|\xi|)}
  \right].
\end{align}
Substituting \eqref{eq:ftrueleft} and \eqref{eq:ftrueright} in
\eqref{eq:ftrue} we have
\begin{equation}
  \label{eq:saepapprox}
  \frac{\dot{\Omega}_{\varrho}(r_{0})}{\Omega_{\varrho}(r_{0})}=
  \eta,
\end{equation}
which determines the parameter $\varrho$ in terms of the physics of the
problem, i.e., the correct behavior of the wave functions for
$r\rightarrow r_{0}$.

Next, we will find the bound states of the Hamiltonian
$H_{0}$ and using \eqref{eq:saepapprox}, the spectrum of
$H$ will be determined without any arbitrary parameter.
Then, from Eq. \eqref{eq:hzeroeigen} we achieve the modified Bessel
equation ($\kappa^{2}=-2M\mathcal{E}$, $\mathcal{E}<0$)
\begin{equation}
\label{eq:eigenvalue}
\left[
  \frac{d^{2}}{dr^{2}}+\frac{1}{r}\frac{d}{dr}-
  \left(\frac{\xi^{2}}{r^{2}}+\kappa^{2}\right)
\right] f_{\varrho,\mathcal{E}}(r) = 0,
\end{equation}
whose general solution is given by
\begin{equation}
  \label{eq:sver}
  f_{\varrho,\mathcal{E}}(r)=K_{|\xi|}
  \left(r\sqrt{-2M\mathcal{E}}\right).
\end{equation}
Since this solutions belongs to $\mathcal{D}(H_{0,\varrho})$,
it is the form \eqref{eq:domain} for some $\varrho$ selected
from the physics of the problem at $r=r_{0}$.
So, we substitute \eqref{eq:sver} in \eqref{eq:domain} and
compute $\dot{f}_{\varrho,\mathcal{E}}/f_{\varrho,\mathcal{E}}\big|_{r=r_{0}}$
using \eqref{eq:besselasympt}.
After a straightforward calculation we have the relation
\begin{align}
  \label{eq:derfe}
  \frac{\dot{f}_{\varrho,\mathcal{E}}(r_0)}{f_{\varrho,\mathcal{E}}(r_0)}
  = {}&
  \frac
  {
    |\xi|
    \left[
      r_{0}^{2|\xi|}(-M \mathcal{E})^{|\xi|}
      +2^{|\xi|} \Theta_{\xi}
    \right]
  }
  {r_{0}^{2|\xi|}(-M \mathcal{E})^{|\xi|}
    -2^{|\xi|} \Theta_{\xi}}
\nonumber \\
  ={}&\frac{\dot{\Omega}_{\varrho}(r_{0})}{\Omega_{\varrho}(r_{0})},
\end{align}
with $\Theta_{\xi}= \Gamma(1+|\xi|)/ \Gamma(1-|\xi|)$.
By using Eq. \eqref{eq:saepapprox} into Eq. \eqref{eq:derfe} and solving
for $\mathcal{E}$, we find the sought energy spectrum
\begin{equation}
  \label{eq:energy_KS}
  \mathcal{E}=
  -\frac{2}{Mr_{0}^{2}}
  \left[
    \left(
      \frac{\eta+|\xi|}
           {\eta-|\xi|}
    \right)
    \frac{\Gamma(1+|\xi|)}{\Gamma(1-|\xi|)}
   \right]^{1/|\xi|}.
\end{equation}
Notice that there are no arbitrary parameters in the above
equation and we must have $\eta\le-1$ to ensure that the energy
is a real number.
The AC problem only has bound states when we take into account the
$\boldsymbol{\nabla}\cdot \mathbf{E}$, which gives an attractive
$\delta$ function.

\section{AC plus a two-dimensional harmonic oscillator}
\label{sec:acho}

In this section, we address the AC system plus a two-dimensional
harmonic oscillator (HO) located at the origin of a polar coordinate
system by using a same approach as in Sec. \ref{sec:bstates_AC}.

The potential of the harmonic oscillator in two-dimensional space is
given by
\begin{equation}
  \label{eq:posc}
  \hat{V}_{HO}=\frac{1}{2}M\omega_{x}^{2}x^{2}+
  \frac{1}{2}M\omega_{y}^{2}y^{2}.
\end{equation}
In polar coordinates $(r,\varphi $) it can be written as
\begin{equation}
\hat{V}_{HO}=\frac{1}{2}M\omega^{2}r^{2},  \label{eq:posp}
\end{equation}
where we considered $\omega_{x}=\omega_{y}=\omega $.

Let us now include the potential of the oscillator \eqref{eq:posp} into
the AC Hamiltonian \eqref{eq:hamil2d}, which leads to the following
eigenvalue equation
\begin{equation}
\hat{H}_{HO}\Phi =i\partial_{t}\Phi ,  \label{eq:eeff}
\end{equation}
where
\begin{equation}
  \hat{H}_{HO}= \hat{H}_{NR}+\hat{V}_{HO}.
\end{equation}
By proceeding in the same manner as in Sec. \ref{sec:bstates_AC}, we can
write the solutions as
\begin{equation}
  \label{eq:soloc}
  \Phi(r,\varphi )=
  \left[
    \begin{array}{c}
      \phi_{\mathcal{E}'}(r)e^{i(m_{j}-1/2)\varphi} \\
      \zeta_{\mathcal{E}'}(r)e^{i(m_{j}+1/2)\varphi}
    \end{array}
\right] ,
\end{equation}
which implies the following radial equation for eigenvalues
\begin{equation}
  \label{eq:eeigen}
  H'\phi_{\mathcal{E}'}(r)=
  \mathcal{E}'\phi_{\mathcal{E}'}(r),
\end{equation}
where
\begin{equation}
H'=H_{0}^{\prime}+\overline{U}_{\rm short},
\end{equation}
\begin{equation}
  \label{eq:hlinha}
  H_{0}^{\prime}=-\frac{1}{2M}
  \left[
    \frac{d^{2}}{dr^{2}}+\frac{1}{r}\frac{d}{dr}
    -\frac{\xi^{2}}{r^{2}}-\gamma^{2}r^{2}
  \right] ,
\end{equation}
and $\gamma^{2}=M^{2}\omega^{2}$.

In order to have a more detailed study of the problem we will analyze
separately the motion of the particle in the $r\neq 0$ region and
including the $r=0$ region.
This approach allows us to see explicitly the
physical implications of the $\boldsymbol{\nabla} \cdot \mathbf{E}$ term
on the energy spectrum of the particle and thus makes it clear that the
fact cannot be neglected in the Hamiltonian.

\subsection{Solution of the problem in the $r\neq 0$ region}
\label{subsec:ACHOrneq0}

In this case, Eq. \eqref{eq:eeigen} does not contain the
$\overline{U}_{\rm short}$ term. Then, the general solution
for  $\phi_{\mathcal{E}'}(r)$ in the $r\neq0$ region is
\cite{Book.2010.NIST}
\begin{align}
  \label{eq:sol1}
  \phi_{\mathcal{E}'}(r)
  = {} &A_{\xi}\gamma^{\frac{1+\xi}{2}}r^{\xi}
  e^{-\frac{1}{2}\gamma r^{2}}M(d,1+\xi,\gamma r^{2})
  \nonumber \\
 &{}+ B_{\xi}\gamma^{\frac{1+\xi}{2}}r^{\xi}
  e^{-\frac{1}{2}\gamma r^{2}}U(d,1+\xi,\gamma r^{2}),
\end{align}
where $d={(1+\xi)}/{2}-{M\mathcal{E}'}/{2\gamma}$,
$M(a,b,z)$ and $U(a,b,z)$ are the confluent hypergeometric functions
(Kummer's functions) \cite{Book.1972.Abramowitz}, $A_{\xi}$ and
$B_{\xi}$ constants.
However, only $M$ is regular at origin, this implies that $B_{\xi}=0$.
In the above solution, moreover, if $d$ is $0$ or
a negative integer, the series terminates and the hypergeometric
function becomes a polynomial of degree $n$
\cite{Book.1972.Abramowitz}.
This condition guarantees that the hypergeometric function is
regular at origin, which is essential for the treatment of the
physical system since the region of interest is that around the
charge filament.
Therefore, the series in \eqref{eq:sol1} must converge if we
consider that $d=-n$, where $n \in\mathbb{Z}^{*}$,
$\mathbb{Z}^{*}$ denoting the set of nonnegative integers,
$n=0,1,2,3,...$.
This condition also guarantees the normalizability of the wave
function. So, using this condition, we  obtain the discrete
values for the energy whose expression is given by
\begin{equation}
\mathcal{E}'=(2n+1+|\xi|)\omega,\qquad n \in \mathbb{Z}^{*}.
\label{eq:en1}
\end{equation}
The energy eigenfunction is given by
\begin{equation}
  \phi_{\mathcal{E}'}(r)=
  C_{\xi}\gamma^{\frac{1+|\xi|}{2}}r^{|\xi|}
  e^{-\frac{1}{2}\gamma r^{2}}
  M(-n,1+|\xi|,\gamma r^{2}),
  \label{eq:wf1}
\end{equation}
where $C_{\xi}$ is a normalization constant.
Notice that in Eq. \eqref{eq:en1}, $|\xi|$ can assume any
noninteger value.
However, we will see that this condition is no longer
satisfied when we include the
$\boldsymbol{\nabla}\cdot \mathbf{E}$ term.
To study the dynamics of the particle in all space, including
the $r=0$ region, we invoke the self-adjoint extension of symmetric
operators.
The procedure used here to derive the result of Eq. \eqref{eq:en1}
is found in many articles in the literature, where the authors simply
ignore the term involving the singularity. As we will show in the next
section, this procedure reflects directly in the energy spectrum of
the system.

\subsection{Solution including the $r=0$ region}
\label{subsec:ACHOreq0}

Now, the dynamics includes the $\overline{U}_{\rm short}$
term. So, let us follow the same procedure as in
Sec. \ref{sec:bstates_AC} to find the bound states for the Hamiltonian
$H'$. Like before we need to find all the self-adjoint
extension for the operator $H_{0}'$. The relevant eigenvalue
equation is
\begin{equation}
  H_{0}'\phi_{\vartheta,\mathcal{E}'}(r)=
  \mathcal{E}'\phi_{\vartheta,\mathcal{E}'}(r),
  \label{eq:autovalor}
\end{equation}
with $\phi_{\vartheta,\mathcal{E}'}$ labeled by a parameter $\vartheta$
which is related to the behavior of the wave function in the limit
$r\to r_{0}$.
The solutions of this equation are given in \eqref{eq:sol1}.
However, the only square integrable function is
$U(d,1+\xi,\gamma r^{2})$.  Then, this implies that
$A_{\xi}=0$ in Eq. \eqref{eq:sol1}, so that
\begin{equation}
  \phi_{\vartheta,\mathcal{E}'}(r)=\gamma^{\frac{1+\xi}{2}}r^{\xi}
  e^{-\frac{1}{2} \gamma r^{2}}
  U(d,1+\xi,\gamma r^{2}).
  \label{eq:newsol}
\end{equation}
To guarantee that $\phi_{\vartheta,\mathcal{E}'}(r)\in L^{2}(\mathbb{R}, rdr)$
it is advisable to study the behavior as $r\rightarrow 0$, which
implies analyzing the possible self-adjoint extensions.

Now, in order to construct the self-adjoint extensions, let us consider
the eigenvalue equation
\begin{equation}
H_{0}'^{\dagger}\phi_{\pm}(r)=\pm i k_{0}\phi_{\pm}(r).
\label{eq:sefa}
\end{equation}
Since $H_{0}'^{\dagger}=H_{0}'$ and, from
\eqref{eq:newsol}, the square integrable solution to above equation is
given by
\begin{equation}
\phi_{\pm}(r)=r^{\xi}e^{-\frac{\gamma r^{2}}{2}}
U(d_{\pm},1+\xi,\gamma r^{2}),
\label{eq:phi}
\end{equation}
where $d_{\pm}={(1+\xi)}/{2}\mp{i k_{0}}/{2\gamma}.$
Let us now consider the asymptotic behavior of
$U(d_{\pm},1+\xi,\gamma r^{2})$ as $r\rightarrow 0$
\cite{Book.2010.NIST},
\begin{equation}
  \label{eq:expan}
  U(d_{\pm},1+\xi,\gamma r^{2})\sim
  \left[
    \frac{\Gamma (\xi)(\gamma r)^{-\xi}}{\Gamma (d_{\pm})}+
    \frac{\Gamma (-\xi)r^{\xi}}{\Gamma (d_{\pm}-\xi)}
  \right].
\end{equation}
Working with the expression, let us find under which condition
\begin{equation}
\int|\phi_{\pm}(r)|^{2}rdr,  \label{eq:intc}
\end{equation}
has a finite contribution from the near origin region. Taking
\eqref{eq:phi} and \eqref{eq:expan} into account, we have
\begin{equation}
  \lim_{r\rightarrow 0}|\phi_{\pm}(r)|^{2}r^{1+2\xi}
  \longrightarrow
  \left[
    \mathcal{A}_{1}r^{1+2\xi}+
    \mathcal{A}_{2}r^{1-2\xi}
\right] ,  \label{eq:lim}
\end{equation}
where $\mathcal{A}_{1}$ and $\mathcal{A}_{2}$ are constants.
The equation above shows that $\phi_{\pm}(r)$ is square integrable
only for $\xi\in(-1,1)$. In this case, since
$\mathcal{N}_{+}$ is expanded by $\phi_{+}(r)$ only, we find that its
dimension $n_{+}=1$. The same applies to $\mathcal{N}_{-}$ and
$\phi_{-}(r)$ resulting in $n_{-}=1$. Then, $H_{0}'$
possesses self-adjoint extensions parametrized by a unitary matrix
$U(1)=e^{i\vartheta}$, with $\vartheta \in [0,2\pi)$.
Therefore, the domain
$\mathcal{D}(H_{0}'^{\dagger})$ in
$L^{2}(\mathbb{R}^{+},r dr)$  is given by
\begin{equation}
\mathcal{D}(H_{0}^{' \dagger})=
\mathcal{D}(H_{0}^{'})\oplus
\mathcal{N}_{+}\oplus \mathcal{N}_{-}.
\label{eq:dom}
\end{equation}
So, to extend the domain $\mathcal{D}(H_{0}')$ to
match $\mathcal{D}(H_{0}^{\prime\dag})$ and therefore make
$H_{0}^{\prime}$ self-adjoint, we get
\begin{equation}
\mathcal{D}(H_{0,\vartheta}^{\prime})=
\mathcal{D}(H_{0}^{\prime\dag})=
\mathcal{D}(H_{0}^{\prime})
\oplus \mathcal{N}_{+}
\oplus \mathcal{N}_{-},
\label{eq:dominio}
\end{equation}
we mean that, for each $\vartheta$, we have a possible domain
for $\mathcal{D}(H_{0,\vartheta}^{\prime})$.
But will be the physical situation which will determine the
value of $\vartheta$.
The Hilbert space \eqref{eq:dominio}, for both cases of
Eq. \eqref{eq:xirange}, contains functions of the form
\begin{align}
  \label{eq:solnew}
  \phi_{\vartheta,\mathcal{E}'}(r)= {} &
  \phi_{|\xi|}(r)+C\;
  r^{|\xi|} e^{-\frac{\gamma \; r^{2}}{2}}
  \Big[
  U(d_{+},1+|\xi|,\gamma r^{2}) \nonumber \\
  {} & +
  e^{i\vartheta}U(d_{-},1+|\xi|,\gamma r^{2})
  \Big],
\end{align}
where $C$ is an arbitrary complex number,
$\phi_{|\xi|}(0)=\dot{\phi}_{|\xi|}(0)=0$ with
$\phi_{|\xi|}(r)\in L^{2}(\mathbb{R}^{+},rdr)$.
For a range of $\vartheta$, the behavior of the wave functions
\eqref{eq:solnew} was addressed in \cite{JMP.1989.30.1053}.
However, as we will see below, if we fix the physics of the
problem at $r=r_{0}$, there is no need for such analysis because
the value of $\vartheta $ is automatically selected.

In order to find the spectrum of $H_{0}'$ we consider
the limit $r\rightarrow 0$ of $\phi_{\vartheta,\mathcal{E}'}(r) $, that
is, we substitute \eqref{eq:newsol} in the left side of
\eqref{eq:solnew} and, using \eqref{eq:expan}, after equating
the coefficients of the same power in $r$, we have
\begin{equation}
  \frac{A}{\Gamma(d)}=C
  \left[
    \frac{1}{\Gamma(d_{+})}+
    \frac{e^{i\vartheta}}{\Gamma(d_{-})}
  \right],
\end{equation}
and
\begin{equation}
  \frac{A}{\Gamma (d-|\xi|)}=C
  \left[
    \frac{1}{\Gamma(d_{+}-|\xi|)}+
    \frac{e^{i\vartheta}}{\Gamma(d_{-}-|\xi|)}
  \right],
\end{equation}
whose quotient leads to
\begin{equation}
  \frac{\Gamma(d-|\xi|)}{\Gamma(d)}=
  \frac
  {\frac{1}{\Gamma(d_{+})}+\frac{e^{i\vartheta}}{\Gamma(d_{-})}}
  {\frac{1}{\Gamma(d_{+}-|\xi|)}
    +\frac{e^{i\vartheta}}{\Gamma(d_{-}-|\xi|)}}.
  \label{eq:ig4}
\end{equation}
The left-hand side of this equation is a function of the energy
$\mathcal{E}'$ while its right-hand side is a constant (even
though it depends on the extension parameter $\vartheta $ which
is fixed by the physics of the problem).
Then we have the equation
\begin{equation}
  \frac
  {\Gamma\left(\frac{1-|\xi|}{2}-\frac{M\mathcal{E}'}{2\gamma}\right)}
  {\Gamma\left(\frac{1+|\xi|}{2}-\frac{M\mathcal{E}'}{2\gamma}\right)}
  ={\rm const.}
\label{eq:espectro}
\end{equation}
Therefore we have achieved the energy levels with an arbitrary
parameter $\vartheta $, and we get inequivalent quantizations to
different values of it \cite{PRA.2007.76.12114}.
Each physical problem selects a specific
$\vartheta $.

Next, we replace problem
$H^{\prime}=H_{0}'+\overline{U}_{\rm short}$,
by $H_{0}'$ plus self-adjoint extensions. So, consider the
static solutions ($\mathcal{E}'=0$) $\phi_{0}(r)$ and
$\phi_{\vartheta,0}(r)$ for the equations
\begin{equation}
  \label{eq:tru1}
  \left[
    H_{0}'+\overline{U}_{\rm short}
  \right]
  \phi_{0}(r)=0,
\end{equation}
\begin{equation}
H_{0}'\phi_{\vartheta,0}(r)=0.  \label{eq:ide1}
\end{equation}
Now, to find the value of $\vartheta$, the continuity of logarithmic
derivative is required
\begin{equation}
  r_{0}\frac{\dot{\phi}_{0}(r_{0})}{\phi_{0}(r_{0})}
  =r_{0}\frac{\dot{\phi}_{\vartheta,0}(r_{0})}{\phi_{\vartheta,0}(r_0)}.
  \label{eq:limig}
\end{equation}
The left-hand side of Eq. \eqref{eq:limig} is obtained by integration of
Eq. \eqref{eq:tru1} from $0$ to $r_{0}$.
Since the integration of the harmonic term
\begin{equation}
  \label{eq:ocom}
  \int_{0}^{r_{0}}\gamma^{2}r^{2}r^{1+2|\xi|}\phi_{0}(r)  dr
  \approx \phi_{0}(r=r_{0})
  \int_{0}^{r_{0}}\gamma^{2}r^{2}r^{1+2|\xi|}dr\rightarrow 0,
\end{equation}
as $r_{0}\rightarrow 0$, the result is the same as in
\eqref{eq:ftrueleft}.
Then, we arrive at
\begin{equation}
    r_{0}\frac{\dot{\phi}_{0}(r_{0})}{\phi_{0}(r_{0})}
  =\eta.
\label{eq:leftderl}
\end{equation}
The right-hand side of Eq. \eqref{eq:limig} is calculated using the
asymptotic behavior of $\phi_{\pm}(r)$ (see Eq. \eqref{eq:expan}) into
Eq. \eqref{eq:solnew}, and it take the form
\begin{equation}
  r_{0}\frac{\dot{\phi}_{\vartheta,0}(r_{0})}{\phi_{\vartheta,0}(r_{0})}=
  \frac{\dot{\Omega}_{\vartheta}'(r_{0})}{\Omega_{\vartheta}'(r_{0})},
\label{eq:rightderl}
\end{equation}
where
\begin{align}
  \label{eq:omegasl}
  \Omega_{\vartheta}'(r) = {} &
  \left[
    \frac{\Gamma(|\xi|)(\gamma r)^{-|\xi|}}{\Gamma(d_{+})}
    + \frac{\Gamma(-|\xi|)r^{|\xi|}}{\Gamma(d_{+}-\xi)}
  \right]
  \nonumber \\
  {} &
  + e^{i\vartheta}
  \left[
    \frac{\Gamma(|\xi|)(\gamma r)^{-|\xi|}}{\Gamma(d_{-})}
    + \frac{\Gamma(-|\xi|)r^{|\xi|}}{\Gamma(d_{-}-|\xi|)}
  \right],
\end{align}
Replacing \eqref{eq:leftderl} and \eqref{eq:rightderl} in
\eqref{eq:limig} we arrive at
\begin{equation}
  \frac{\dot{\Omega}_{\vartheta}'(r_{0})}{\Omega_{\vartheta}'(r_{0})}
  =\eta.
  \label{eq:fim}
\end{equation}
With this relation we select an approximated value for parameter
$\vartheta$ in terms of the physics of the problem.
The solutions for \eqref{eq:autovalor} are given by \eqref{eq:newsol}
and since this function belongs to
$\mathcal{D}(H_{0,\vartheta}')$, it is of the form
\eqref{eq:solnew} for some $\vartheta$ selected from the physics
of the problem at $r=r_{0}$.
So, using \eqref{eq:newsol} in \eqref{eq:solnew} we have
\begin{align}
  \label{eq:derfel}
  r_{0}
  \frac
  {\dot{\phi}_{\vartheta,\mathcal{E}'}(r_{0})}
  {\phi_{\vartheta,\mathcal{E}'}(r_{0})}
  = {} &
  \frac
  {|\xi| [r_{0}^{2|\xi|}\gamma^{|\xi|} \Gamma(d)+
    \Theta_{\xi} \Gamma(d-|\xi|)]}
  {r_{0}^{2 |\xi|} \gamma^{|\xi|} \Gamma(d)-
   \Theta_{\xi} \Gamma (d-|\xi|)}
 \nonumber \\
  ={} &\frac{\dot{\Omega}_{\vartheta}'(r_{0})}{\Omega_{\vartheta}'(r_{0})}.
\end{align}
Using Eq. \eqref{eq:fim} into the above equation we obtain
\begin{equation}
\label{eq:energy_KS_HO}
  \frac
  {\Gamma(\frac{1+|\xi|}{2}-\frac{M\mathcal{E}'}{2\gamma})}
  {\Gamma(\frac{1-|\xi|}{2}-\frac{M\mathcal{E}'}{2\gamma})}=
  -\frac{1}{\gamma^{|\xi|} r_{0}^{2|\xi|}}
  \left(
    \frac
    {\eta+|\xi|}
    {\eta-|\xi|}
  \right)
    \frac
  {\Gamma(1+|\xi|)}
  {\Gamma(1-|\xi|)}.
\end{equation}
Eq. \eqref{eq:energy_KS_HO} is too complicated to evaluate
the bound state energy explicitly, but its limiting features are
interesting.
If we take the limit $r_{0}\to 0$ in this expression, the bound
state energy is determined by the poles of the gamma functions,
i.e.,
\begin{equation}
  \mathcal{E}'=
  \left\{
    \begin{array}{llr}
      (2n+1-|\xi|)\omega, &\mbox{for } & -1<\xi<0, \\
      (2n+1+|\xi|)\omega, &\mbox{for } &  0<\xi<1,
    \end{array}
  \right.
\end{equation}
or
\begin{equation}
\mathcal{E}'=(2n+1\pm|\xi|) \omega ,
\label{eq:landau+}
\end{equation}
where $n\in \mathbb{Z}^{*}$. The $+$ ($-$) sign refers to solutions
which are regular (singular) at the origin. This result coincides with
the study done by Blum \textit{et al.} \cite{PRL.1990.64.709}.
It should be noted that in the absence of spin (i.e., when
$\boldsymbol{\nabla}\cdot \mathbf{E}$ is absent), like showed in
Sec. \ref{subsec:ACHOrneq0}, we always have a regular solution
it is the plus sign which must be used in \eqref{eq:landau+}.
In the next section we will confirm this using another approach.

Another interesting case is that of vanishing harmonic oscillator
potential. This is achieved using the asymptotic behavior of the ratio
of gamma functions for $\gamma\to 0$ \cite{Book.2010.NIST},
\begin{equation}
  \frac
  {\Gamma\left(\frac{1+|\xi|}{2}-\frac{M\mathcal{E}}{2\gamma}\right)}
  {\Gamma\left(\frac{1-|\xi|}{2}-\frac{M\mathcal{E}}{2\gamma}\right)}
  \sim
  \left(\frac{-M\mathcal{E}}{2\gamma}\right)^{|\xi|},
\label{eq:omegalimit}
\end{equation}
which holds for $\mathcal{E}<0$ and this condition is necessary
for the usual AC system to have a bound state.
Using this limit in Eq. \eqref{eq:energy_KS_HO} one finds
\begin{equation}
  \label{eq:aclimit}
  \left(\frac{-M\mathcal{E}}{2\gamma}\right)^{|\xi|}=
  -\frac{1}{\gamma^{|\xi|} r_{0}^{2|\xi|}}
  \left(
    \frac
    {\eta+|\xi|}
    {\eta-|\xi|}
  \right)
    \frac
  {\Gamma(1+|\xi|)}
  {\Gamma(1-|\xi|)}.
\end{equation}
Then, by solving Eq. \eqref{eq:aclimit} for $\mathcal{E}$, we
obtain
\begin{equation}
  \label{eq:energy_pure_AC}
  \mathcal{E}=
  -\frac{2}{Mr_{0}^{2}}
  \left[
    \left(
      \frac{\eta+|\xi|}
           {\eta-|\xi|}
    \right)
    \frac{\Gamma(1+|\xi|)}{\Gamma(1-|\xi|)}
   \right]^{1/|\xi|}.
\end{equation}
which agrees with the result obtained in
Eq. \eqref{eq:energy_KS}.
Thus, in the limit of vanishing harmonic oscillator, we recover
the usual AC problem.

Now we have to remark that this result contains a subtlety that
must be interpreted as follows: the presence of the singularity
restricts the range of $\xi$ given by $\xi\in(-1,1)$.
If we ignore the singularity and impose that the wave function
be regular at the origin ($\phi(0)\equiv\dot{\phi}(0)\equiv0$), 
we achieve an incomplete answer to spectrum
\eqref{eq:landau+}, because only the plus sign is present in
the equation and $\xi$ can have any noninteger value.
\cite{PLA.1994.195.90,JPA.2000.33.5513,EPL.1999.45.279}.
In this sense, the self-adjoint extension approach prevents us
from obtaining a spectrum incompatible with the singular nature
of the Hamiltonian at hand when we have \eqref{eq:ushort_r0}
\cite{PRA.2008.77.036101,PRD.1996.53.6829}.
We must to take into account that the true boundary condition is
that the wave function must be square integrable through all
space and it does not matter if it is singular or not at the
origin \cite{CMP.1991.139.103,PRD.1996.53.6829}.

\section{Determination of self-adjoint extension parameter}
\label{sec:dsaep}

Following the procedure describe in \cite{PRD.2012.85.041701}, here we
determine the so called self-adjoint extension parameter for the AC
problem plus a harmonic oscillator and for the usual AC
system.
The approach used in the previous sections gives us the energy spectrum
in terms of the physics of the problem, but is not appropriate for
dealing with scattering problems.
Furthermore, it selects the values to parameters $\varrho$ and
$\vartheta$.
On the other hand, the approach in \cite{JMP.1985.26.2520} is suitable
to address both bound and scattering scenarios, with the disadvantage of
allowing arbitrary self-adjoint extension parameters.
By comparing the results of these two approaches for bound states, the
self-adjoint extension parameter can be determined in terms of the
physics of the problem.
Here, all self-adjoint extensions $H_{0,\alpha_{\xi}}'$  of
$H_{0}'$ are parametrized by the boundary condition at the
origin,
\begin{equation}
  \lim_{r\rightarrow 0^{+}}r^{|\xi|} g(r)=
  \alpha_{\xi}\lim_{r\rightarrow 0^{+}}\frac{1}{r^{|\xi|}}
  \left[
    g(r)-
    \left[
      \lim_{r' \rightarrow 0^{+}}{r'}^{|\xi|} g(r')
    \right]
    \frac{1}{r^{|\xi|}}
  \right],
  \label{eq:bc}
\end{equation}
where $\alpha_{\xi}$ is the self-adjoint extension parameter. In
\cite{Book.2004.Albeverio} is showed that there is a relation between
the self-adjoint extension parameter $\alpha_{\xi}$ used here, and the
parameters $\varrho$ and $\vartheta$  used in the previous sections.
The parameters $\varrho$ and $\vartheta$, which are associated with the
mapping of deficiency subspaces, that extend the domain of operator to
make it self-adjoint, are mathematical parameters.
The self-adjoint extension parameter $\alpha_{\xi}$ has a physical
interpretation, it represents the scattering length
\cite{Book.2011.Sakurai} of $H_{0,\alpha_{\xi}}$
\cite{Book.2004.Albeverio}.
For  $\alpha_{\xi}=0$ we have the free Hamiltonian (without the
$\delta$ function)  with regular wave functions at origin and for
$\alpha_{\xi}\neq 0$  the boundary condition in \eqref{eq:bc} permit a
$r^{-|\xi|}$ singularity in the wave functions at origin.

For our intent, it is more  convenient to write the solutions for
\eqref{eq:eeigen}  for  $r\neq 0$, taking into account both cases in
\eqref{eq:xirange} simultaneously, as
\begin{align}
  \label{eq:general_sol}
  \phi_{\mathcal{E}'}(r)
  ={} &A_{\xi}\gamma^{\frac{1+|\xi|}{2}}
  e^{-\frac{\gamma r^{2}}{2}}r^{|\xi|}
  M(d,1+|\xi|,\gamma r^{2})
  \nonumber \\
  {} &+B_{\xi}\gamma^{\frac{1-|\xi|}{2}}
  e^{-\frac{\gamma r^{2}}{2}}r^{-|\xi|}
  M(d-|\xi|, 1-|\xi|,\gamma r^{2}),
\end{align}
where $A_{\xi}$, $B_{\xi}$ are the coefficients of
the regular and singular solutions, respectively. By implementing
Eq. \eqref{eq:general_sol} into the boundary condition \eqref{eq:bc},
we derive the following relation between the coefficients
$A_{\xi}$ and $B_{\xi}$:
\begin{equation}
  \alpha_{\xi}' A_{\xi}\gamma^{|\xi|}=B_{\xi}
  \bigg(
  1-\frac{\alpha_{\xi}'\;\mathcal{E}'}{4(1-|\xi|)}
  \lim_{r\rightarrow 0^{+}}r^{2-2|\xi|}
  \bigg),  \label{eq:coef_rel_1}
\end{equation}
where $\alpha_{\xi}'$ is the self-adjoint extension parameter for
the AC plus HO system. In the above equation, the coefficient of
$B_{\xi}$ diverges as $\lim_{r\rightarrow 0^{+}}r^{2-2|\xi|}$, if
$|\xi|>1.$ Thus, $B_{\xi}$
must be zero for $|\xi|>1$, and the condition for the occurrence of a
singular solution is $|\xi|<1$. So, the presence of an irregular
solution stems from the fact the operator is not self-adjoint for
$|\xi|<1$, recasting the condition of non-self-adjointness of the
previews sections,  and this irregular solution is associated with a
self-adjoint extension of the operator $H_{0}'$
\cite{JPA.1995.28.2359,PRA.1992.46.6052}.
In other words, the self-adjoint extension essentially consists in
including irregular solutions in $\mathcal{D}(H_{0}')$ to
match $\mathcal{D}(H_{0}'^{\dagger})$, which allows us to
select an appropriate boundary condition for the problem.

In order to Eq. \eqref{eq:general_sol} to be a bound state,
$\phi_{\xi,\mathcal{E}'}(r)$ must vanish for large values of $r$,
i.e., must be normalizable at large $r$. By using the asymptotic
representation of $M(a,b,z)$ for $z\rightarrow \infty$,
\begin{equation}
  M(a,b,z) \to
  \frac{\Gamma(b)}{\Gamma(a)} e^{z} z^{a-b} +
  \frac{\Gamma(b)}{\Gamma(b-a)}(-z)^{-a},
\end{equation}
the normalizability condition yields the relation
\begin{equation}
  B_{\xi}=
  -\frac
  {\Gamma(1+|\xi|)}
  {\Gamma(1-|\xi|)}
  \frac
  {\Gamma\left(\frac{1+|\xi|}{2}-\frac{\mathcal{E}'}{2\gamma}\right)}
  {\Gamma\left(\frac{1-|\xi|}{2}-\frac{\mathcal{E}'}{2\gamma}\right)}
  A_{\xi}.
\label{eq:coef_rel_2}
\end{equation}
From Eq. \eqref{eq:coef_rel_1}, for $|\xi|<1$ we have
$B_{\xi}=\alpha_{\xi}'\gamma^{|\xi|}A_{\xi}$ and by using
Eq. \eqref{eq:coef_rel_2}, the bound state energy is implicitly
determined by the equation
\begin{equation}
  \frac
  {\Gamma\left(\frac{1+|\xi|}{2}-\frac{M\mathcal{E}'}{2\gamma}\right)}
  {\Gamma\left(\frac{1-|\xi|}{2}-\frac{M\mathcal{E}'}{2\gamma}\right)}=
  -\frac{1}{\alpha_{\xi}'\gamma^{|\xi|}}
  \frac
  {\Gamma(1+|\xi|)}
  {\Gamma(1-|\xi|)}.
  \label{eq:energy_BG}
\end{equation}
By comparing Eq. \eqref{eq:energy_BG} with Eq. \eqref{eq:energy_KS_HO},
we find
\begin{equation}
  \frac{1}{\alpha_{\xi}'}=\frac{2}{r_{0}^{2|\xi|}}
  \left(
    \frac
    {\eta+|\xi|}
    {\eta-|\xi|}
  \right).
  \label{eq:alpha_xi}
\end{equation}
We have thus attained a relation between the self-adjoint extension
parameter and the physical parameters of the problem.
Eq. \eqref{eq:energy_BG} coincides with Eq. (53) of
Ref. \cite{JMP.1995.36.5453} for the AB system with the exact
equivalence condition between the vector potential and electric field
\cite{PRL.1990.64.2347}, i.e.,
\begin{equation}
eA_{i}=\mu s \epsilon_{ij}E_{j}.
\end{equation}
The limiting features of Eq. \eqref{eq:alpha_xi} are interesting.
For $\alpha_{\xi} \to 0$  or $\infty$, from the poles of gamma function,
we have
\begin{equation}
  \mathcal{E}'=
  \left\{
    \begin{array}{lll}
      (2n+1+|\xi|)\omega, &\mbox{for } & \alpha_{\xi}'=0, \\
      (2n+1-|\xi|)\omega, &\mbox{for } & \alpha_{\xi}'=\infty.
    \end{array}
  \right.
\end{equation}
These bound states energies coincide with those regular and irregular
solutions given in Eq. \eqref{eq:landau+} of the previous section.
From relation \eqref{eq:alpha_xi} the regular wave function, when
$\alpha_{\xi}'=0$ and the $\boldsymbol{\nabla}\cdot \mathbf{E}$ is
absent, is associated with $0<\xi<1$, and the irregular wave
function, when $\alpha_{\xi}'=\infty$, is associated with $-1<\xi<0$.

Like before, another interesting limit is that of the vanishing
oscillator potential. So, using the limit \eqref{eq:lim} in
\eqref{eq:energy_BG}, we have
\begin{equation}
  \label{eq:aclimit_BG}
  \frac{1}{2^{|\xi|}}
  \left(\frac{-M\mathcal{E}}{\gamma}\right)^{|\xi|}=
  -\frac{1}{\alpha_{\xi} \gamma^{|\xi|}}
  \frac
  {\Gamma(1+|\xi|)}
  {\Gamma(1-|\xi|)},
\end{equation}
with $\alpha_{\xi}$ the self-adjoint extension for the usual AC system.
From this equation we have the spectrum for the usual AC system in terms
of the self-adjoint extension parameter,
\begin{equation}
  \mathcal{E}=
  -\frac{2}{M}
  \left[
    -\frac{1}{\alpha_{\xi}}
    \frac
    {\Gamma(1+|\xi|)}
    {\Gamma(1-|\xi|)}
   \right]^{1/|\xi|}.
  \label{eq:energy_BG_pure}
\end{equation}
This result also coincides with Eq. (3.13) of
Ref. \cite{PRD.1994.50.7715} for the AB system, with the exact
equivalence cited above.
By comparing this equation with \eqref{eq:energy_pure_AC} we
arrive at
\begin{equation}
  \frac{1}{\alpha_{\xi}}=-\frac{2}{r_{0}^{2|\xi|}}
  \left(
    \frac
    {\eta+|\xi|}
    {\eta-|\xi|}
  \right),
\end{equation}
for the relation of the self-adjoint extension parameter and the physics
of the problem for usual AC system. Then, the relation between the
self-adjoint extension parameter and the physics of the problem for the
usual AC has the same mathematical structure as for the AC plus
HO. However, we must observe that the self-adjoint extension parameter
is negative for the usual AC, confirming the restriction of negative
values of the self-adjoint extension made in \cite{JMP.1995.36.5453}, in
such way we have an attractive $\delta$ function. It is a necessary
condition to have a bound state in the usual AC system.

It should be mentioned that some relations involving the self-adjoint
extension parameter and the $\delta$ function coupling constant (here
represented by $\eta$) were previously obtained by using Green's
function in \cite{JMP.1995.36.5453} and renormalization technique in
\cite{Book.1995.Jackiw}, both, however, lacking a clear physical
interpretation.

\section{Conclusions}
\label{sec:conclusions}

By modeling the problem by boundary conditions at origin which arises by
the self-adjoint extension of the nonrelativistic Hamiltonian, we have
presented, for the first time, an expression for the energy spectrum of
a spin-1/2 neutral particle with a magnetic moment $\mu$ moving in a
plane subject to an electric field, i.e., for the usual AC system taking
into account the $\boldsymbol{\nabla}\cdot \mathbf{E}$ term,  which
features a point interaction between the particle and the charged line.
The presence of the $\boldsymbol{\nabla}\cdot\mathbf{E}$ term,  which
gives rise to an attractive $\delta$ function for $\lambda<0$,
ensures that we have at least one bound state.

As an application, we also addressed the AC problem plus a
two-di\-men\-sion\-al
harmonic oscillator located at the origin of a polar coordinate system
by including the term $M\omega^{2}r^{2}/2$ in the
nonrelativistic Hamiltonian.
Two cases were considered: (i) without and (ii) with the inclusion of
$\boldsymbol{\nabla}\cdot \mathbf{E}$ in the nonrelativistic
Hamiltonian.
Even though we have obtained an equivalent mathematical expression for
both cases, it has been shown that in (i) $\xi$ can assume any
value while in (ii) it is in the range $\xi\in(-1,1)$.
In the first case, it is reasonable to impose that the wave function
vanish at the origin. However, this condition does not give a correct
description of the problem in the $r=0$ region.
Therefore, the energy spectrum obtained in the second case is
physically acceptable.

Finally, we determine the self-adjoint extension parameter
for the AC plus HO and for the usual AC problem based on the physics of
the problem. Like showed in \cite{PRD.2012.85.041701} this
self-adjoint extension parameter can be used to study scattering. The
scattering in the AC system was studied in
\cite{PRL.1990.64.2347,MPLA.2010.25.1531,JPA.2010.43.354008} using
other methods than self-adjoint extension.
Results in this subject using self-adjoint extensions
will be reported elsewhere.

\section*{Acknowledgements}
The authors would like to thank F. Moraes for their critical
reading of the manuscript, and for helpful discussions.
E. O. Silva acknowledges research grants by CNPq-(Universal)
project No. 484959/2011-5,
H. Belich and C. Filgueiras acknowledges research grants by CNPq
(Brazilian agencies).

\bibliographystyle{spphys}

\end{document}